\documentclass{article}
\begin{document}
\begin{center}
\textbf{Reply to the comment of Chudnowski and Garanin on
Phonon Assited Tunneling in Mn$_{12}$ (PRL 83, 416 (1999))}\\
\vspace{0.5cm}
G. Bellessa$^{1}$, N. Vernier$^{1}$, B. Barbara$^{2}$ 
and D. Gatteschi$^{3}$\\
\end{center}
\textit{$^1$ Laboratoire de Physique des Solides, B\^atiment 510, Universit\'e 
Paris-Sud, 91405 Orsay, France}\\
\textit{$^2$ Laboratoire de Magn\'etisme Louis N\'eel, CNRS, BP 166, 
38042 Grenoble, France}\\
\textit{$^3$ Dipartimento di Chimica, Universit\`a di Firenze, Via  
Maragliano 77, 50144 Firenze, Italia}\\
\vspace{0.5cm}

Chudnovsky  and Garanin have attacked harshly our paper ~\cite{bellessa} in 
a comment in arXiv:cond-mat~\cite{chudnovski}. They write in their comment 
that ``all formulas of Ref. [1] are incorrect''. They question `` how can 
$\Delta$ be the magnetic dipole matrix element and the tunneling splitting
at the same time?''. These two $\Delta$ are obviously different and we write
it explicitlely in the text:
\begin{itemize}
\item The first $\Delta$ appears in the expression of the fundamental doublet:
\begin{equation}                
\sqrt{\Delta^2 + \epsilon^2}
\label{delta}
\end{equation}
$\Delta$ is called the tunneling rate and $\epsilon$ is a Zeeman term. We 
seize the opportunity of restablishing the historical truth of  
Eq.~\ref{delta}. It is Korenblit and Shender who the first have pointed out
that high spins with large anisotropy give tunneling states, have 
resolved analytically the Hamiltonian in the high-spin approximation and have 
given the explicite form of $\Delta$ in Eq.~\ref{delta} \cite{KS}. This is 
often ignored by the MQT community.
\item In our paper there is a second $\Delta$ which is obviously different 
from the first one. We write (Eq.~3 of our paper):
\begin{equation}
\chi'' = C N \Delta^{2} T_{2} tanh(\frac{\hbar \omega}{2 k_{B} T})
\label{chi}
\end{equation} 
and then we write:``where $\Delta$ is the magnetic dipole matrix element 
betwen the two states of the fundamental doublet''. We take the same 
definition as Abragam and Bleaney in their formula that has been called 
`` trivial'' by Chudnowsky and Garanin~\cite{AB}. Hence, this $\Delta$ is 
explicitly different from the first one which is the splitting (equal to 
$\hbar \omega_{0}$, where $\omega_{0}$ is the resonance frequency and
depends on the applied magnetic field).
\end{itemize}
Another way to show that the two $\Delta$ are different is to treat the 
fundamental doublet as an effective spin $S = 1/2$ and   to write the 
Hamiltonian:
\begin{equation}
\mathcal{H} = \mathcal{H}_0 + \mathcal{H}'
\label{hamilt}
\end{equation}
where $\mathcal{H}_0$ is the unperturbed hamiltonian in the representation of 
the unperturbated eigenstates $|1>$ and $|2>$. The eigenstates are coupled to 
the electromagnetic field, which gives $\mathcal{H}$'.  
In our case the magnetic field is perpendicular to anisotropy axis. So, the 
Zeeman term is zero and the splitting $E = \sqrt{\Delta^2 + \epsilon^2}$ is 
$\Delta$. Then the Hamiltonian is:
\begin{equation}
\mathcal{H} = 1/2 \left( \begin{array}{cc} \Delta & 0 \\ 0 & -\Delta \\ 
\end{array} \right) + 1/2 \left( \begin{array}{cc} \delta & M \\
M & -\delta\\ \end{array} \right)
\label{hamilt2}
\end{equation}
where $M$ is the matrix element $<-|\mathcal{H}'|+>$ and $\delta$ is 
negligeable as compared with $\Delta$. The second $\Delta$ in our paper is 
now $M$. It is this matrix element which induces absorption (and/or emission) 
of electromagnetic quanta and transitions within the fundamental doublet, 
these transitions being obviously tunneling transitions.

Eq.~\ref{chi} gives a susceptibility $\chi''$ constant at low  
temperature ($\hbar \omega < k_B T$). However, we observe a strong increase 
of $\chi''$ with increasing temperature. The only parameter in 
Eq.~\ref{chi} is $\Delta$ (the second one of our paper or M now), which 
means that this one increases with temperature, i.e. when phonons are added.  
This is an experimental fact.

To conclude, the attack of Chudnovski and Garanin upon the tunneling rate
in $Mn_{12}$ is absolutely irrelevant.

\end{document}